\documentclass[11pt,reqno,twocolumn]{article}
\usepackage[usenames]{color}
\usepackage[colorlinks=true,
linkcolor=webgreen,
filecolor=webbrown,
citecolor=webgreen]{hyperref}
\definecolor{webgreen}{rgb}{0,.5,0}
\definecolor{webbrown}{rgb}{.6,0,0}

\usepackage{epsfig}                                                                               
\usepackage{amsmath,amssymb}
\usepackage{amsfonts}
\usepackage{verbatim} 
\newcommand{\ud}{\mathop{}\!\mathrm{d}}

\renewcommand{\v}[1]{\mathbf{#1}}
\newcommand{\grad}{\nabla}
\newcommand{\efold}{\epsilon}

\newcommand{\diB}{\Phi} 

\begin{document}
\title{Fringe field simulations of a non-scaling FFAG accelerator}
\author{George~I.~Bell and Dan~T.~Abell \\Tech-X Corporation \\5621 Arapahoe Ave. \\Boulder CO 80303 \\gibell@txcorp.com}


\twocolumn[
\begin{@twocolumnfalse}
\maketitle
\begin{abstract}
Fixed-field Alternating Gradient (FFAG) accelerators
offer the potential of high-quality, moderate energy ion beams at low cost.
Modeling of these structures is challenging with conventional
beam tracking codes because of the large radial excursions of the
beam and the significance of fringe field effects.
Numerous tune resonances are crossed during the acceleration,
which would lead to beam instability and loss in a storage ring.
In a non-scaling FFAG, the hope is that these resonances can be crossed
sufficiently rapidly to prevent beam loss.
Simulations are required to see if this is indeed the case.
Here we simulate a non-scaling FFAG
which accelerates protons from 31 to 250~MeV.
We assume only that the bending magnets have mid-plane symmetry,
with specified vertical bending field in the mid-plane ($y=0$).
The magnetic field can be obtained everywhere using a power series expansion,
and we develop mathematical tools for calculating this expansion to arbitrary order
when the longitudinal field profile is given by an Enge function.
We compare the use of a conventional
hard-edge fringe with a more accurate, soft-edge fringe field model.
The tune 1/3 resonance is the strongest, and crossing
it in the hard-edge fringe model results in a 21\% loss of the beam.
Using the soft-edge fringe model the beam loss is less than 6\%.
\end{abstract}
\end{@twocolumnfalse}
]

\section{Introduction}
The concept of a fixed-field, alternating gradient (FFAG) accelerator
came to light in the 1950's \cite{FFAG1953, FFAG1956, FFAG1956a},
but has only recently been revived as a promising design for a low-cost,
reliable accelerator \cite{Craddock2005, Mori2005}.
Reasons for a lower cost are the fixed-field magnets (as opposed to those that must be ramped up during each
acceleration cycle), small size, and only two different magnet types.

FFAG accelerators come in two varieties: scaling and non-scaling.
In a scaling FFAG the magnets are designed so that the vertical and horizontal
tunes are constant over the acceleration phase.
This guarantees stability of the particle orbits, but at the expense of more complex
magnets which can be difficult to manufacture.
Another problem is the large radial excursion that particles undergo from the lowest to the highest energy,
making the machines (and magnets) larger.
 
Here we are interested instead in a non-scaling FFAG---in such a machine
no attempt is made to keep the tunes constant during the acceleration.
In fact, the tunes vary considerably during the acceleration, crossing many
resonances \cite{Lee2006, Lee2006a, Machida2008}.
The magnets in these devices can be simpler, and the radial excursion of the particles less.
Simulation of these devices is critical and inherently non-linear.
The success of the machine rests on the acceleration being rapid enough that the
instabilities do not have sufficient time to destabilize the beam.

Non-scaling FFAGs have been proposed for use in a muon accelerator \cite{Johnstone1999, Tr2005, Berg2003}
as well as for hadron therapy \cite{Keil2007, Tr2007}.
The world's first non-scaling FFAG, EMMA \cite{Clery2010, Barlow2010}
will allow scientists to study the unique beam dynamics of this kind of machine.

\section{A sample FFAG}

We consider a simple FFAG for proton acceleration consisting of 24 doublet sections.
Each doublet consists of a pair of combined-function bend magnets (``cf bend") with parameters given in
Table~\ref{table0}.
We consider the magnetic field $\v{B}(x,y,z)=(b_x,b_y,b_z)$ where the vertical component $b_y$
within the magnet has the form
\begin{eqnarray}
b_y(x,0,z) = (B_0+Gx)S(z) ,
\label{eq:lingrad}
\end{eqnarray}
in other words a vertical bending field $B_0$
with a linear transverse gradient $G$.
In the vertical coordinate $y$ the magnet is assumed to have mid-plane symmetry.
Mid-plane symmetry will be explained in detail in the next section,
and we will generalize the transverse profile beyond linear gradients.
The function $S(z)$ describes the longitudinal profile of the field.
It is approximately one inside the bending magnet, and decreases smoothly to zero outside it.

\begin{table}[!t]
  \renewcommand{\arraystretch}{1.3}
  \caption{\label{table0}Design parameters for a sample FFAG.}
  \centering
  \begin{tabular}{ccccc} \hline
    \footnotesize{Element} & Type & l (cm) & $B_0$ (T) & \multicolumn{1}{c}{$G$~(T/m)} \\
    \hline
    DL & drift   & 40  &          &       \\
    DS & drift   & 7.5 &          &       \\
    BD & cf bend & 22  & 0.803952 & -12.8 \\
    BF & cf bend & 44  & 0.555057 &   8.0 \\
    \hline
  \end{tabular}
\end{table}

\begin{figure}[htb]
\center
\includegraphics{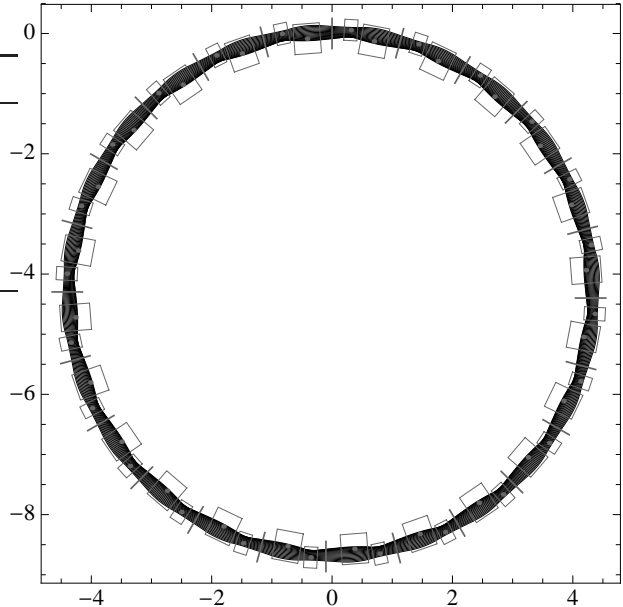}
\caption{Top view of the FFAG ring, showing magnets and closed orbits from 31 to 250~MeV. The scale is in meters.}
\label{figtr2}
\end{figure}

The transverse gradient $G$ is large enough
that the vertical bending field switches sign within the magnet.
This is necessary to keep the increase in overall radius small over an order of magnitude increase in energy
(Figure~\ref{figtr2}).

\begin{figure}[htb]
\center
\includegraphics{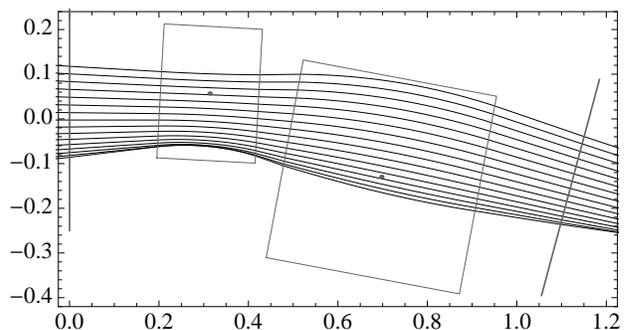}
\caption{Top view of one period in the sample FFAG ring, showing closed orbits from 31 to 250~MeV. The scale is in meters.}
\label{figtr1}
\end{figure}

These combined function magnets can also be considered as quadrupoles which
have been shifted transversely.  
The field in the geometrical center of these magnets is not zero unlike a conventional quadrupole.
In this sense they are a pair of focusing and defocusing quadrupoles.
Figure~\ref{figtr1} shows a single period of the lattice, with the small defocusing quadrupoles
and larger focusing quadrupoles.
These magnets are shown off center because they are conventional quadrupoles shifted off the beam-line.
The dots in the centers of these magnets are the magnetic origins of these magnets, where the field is zero.

This design allows for a large energy acceptance, from 31~MeV to 250~MeV protons,
while the radius of the associated closed orbits ranges from 4~m to 4.2~m.
Figure~\ref{figtr2} shows closed orbits of 16 particles with energies ranging from 31 to 250~MeV.

\begin{figure}[htb]
\center
\includegraphics[scale=0.90]{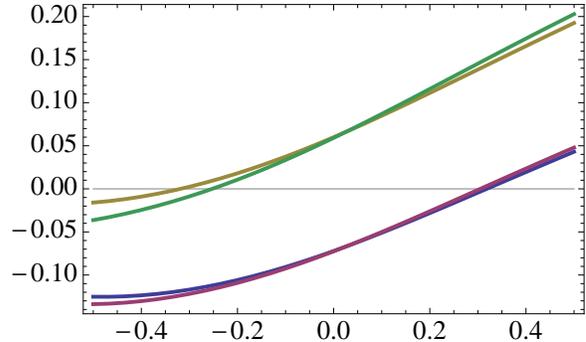}
\caption{Horizontal displacement in meters versus momentum deviation, at the entrance and exit to
the defocusing quad (blue and red), and the focusing quad (tan and green).
Note that 20~cm deviations occur.}
\label{figdx}
\end{figure}

Modeling this accelerator through the entire energy range is a challenge for any tracking code.
The closed orbits range outward nearly 20~cm,
so at the extremes of energy they are far from the center of the magnet.
In Figures~\ref{figtr2} and \ref{figtr1},
fringe fields are modeled in the conventional way by giving each particle a position and momentum kick
as it enters or exits a magnet.
The position kicks contain terms proportional to the position of the particle cubed \cite{Forest88},
which can become quite significant
for a particle displaced 0.2~m from the center of the magnet (Figure~\ref{figdx}).
In this paper we compare the effects of such a ``hard-edge fringe" model with a more accurate field model where the
rapid rise of the magnetic field in the fringe field region is directly integrated through.

\begin{figure}[htb]
\center
\includegraphics[scale=0.90]{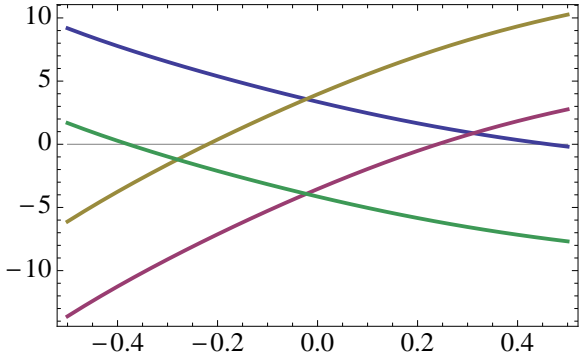}
\caption{Horizontal entrance and exit angles (in degrees) versus momentum deviation for
the defocusing quad (blue and red), and the focusing quad (tan and green).}
\label{figang}
\end{figure}

Figure~\ref{figang} shows the entrance and exit angles to the quads over the total range in energy.
In the fringe field region, when these angles are large they provide significant vertical focusing.

\section{Magnets with mid-plane symmetry}

We consider a rectangular bend magnet in Cartesian coordinates.
The primary bending field is oriented along the $y$-axis,
so that the bending takes place in the $x$-$z$ plane.
The primary bending magnetic field $\v{B}(x,y,z)=(b_x,b_y,b_z)$
is $B_0\hat{y}$ in the center of the bending magnet.
This magnet is assumed to have a specified vertical field in the mid-plane ($y=0$),
\begin{eqnarray}
b_y(x,0,z) = T(x)S(z) ,
\end{eqnarray}
where $T(x)$ is the transverse field profile and
$S(z)$ is a smooth function describing the longitudinal variation of the bending field
through the magnet.
We often will take the transverse field profile to be linear as in
(\ref{eq:lingrad}) but the formulas below give the general case.
The longitudinal profile function $S(z)$ is close to one within the magnet and
decreases to zero outside it.
$S(z)$ can be determined by curve-fitting the field of a real magnet.
A general function with this shape can be obtained using a
$k$-parameter Enge function \cite{Bertz2000, Makino2009},
\begin{eqnarray}
S(z;\efold) & = & \frac{1}{1+\exp\left(2p\left(z/\efold\right)\right)} \nonumber \\
 & = & \frac{1}{2}\left(1-\tanh\left(p(z/\efold)\right)\right)
\label{eq:defenge}
\end{eqnarray}
where $p(z)$ is a $k$ term polynomial of degree $k-1$,
\begin{equation}
p(z)=a_1+a_2 z+a_3 z^2+\ldots+a_k z^{k-1} ,
\end{equation}
and $\efold$ is a length scale over which the longitudinal field changes in the fringe region.
Physically $\efold$ is determined by the aperture of the magnet.
Note that $0<S(z)<1$.
We assume that $k$ is even and that the highest order coefficient $a_k>0$,
so that $S(z)\rightarrow 1$ as $z\rightarrow -\infty$ and
$S(z)\rightarrow 0$ as $z\rightarrow +\infty$.

\begin{figure}[htb]
\center
\includegraphics[width=3.4in]{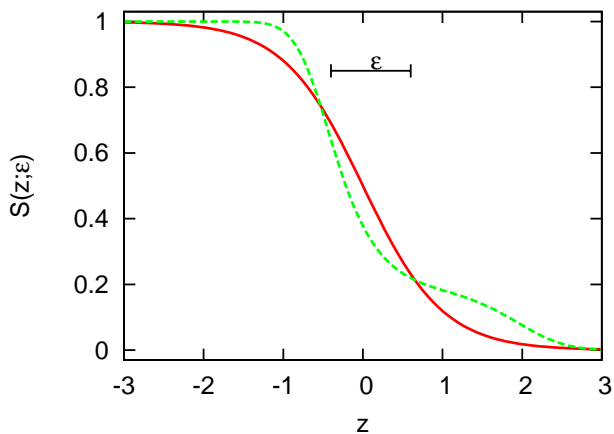}
\caption{Sample longitudinal profile functions $S(z;\efold)$, with $p(z)=z$ (hyperbolic tangent, solid)
as well as $p(z)=(1+4x-3x^2+x^3)/4$ (dashed).  The scaling factor $\efold$ was set to $1$.}
\label{figprofile}
\end{figure}

By adding two Enge functions, we can create a profile function of a magnet
from $z=a$ to $b$
\begin{equation}
\mathcal{S}(z) = -S(z-a;\efold_a) + S(z-b;\efold_b) .
\label{eq:prof}
\end{equation}
The locations $a$ and $b$ correspond to the physical entrance and exit
of the magnet.
Note that the hyperbolic tangent profile is a special case of the 2-parameter Enge function
with $p(z)=z$.
Two Enge functions are shown in Figure~\ref{figprofile}, first the hyperbolic tangent
and then a more general 4-parameter Enge function.

The two profile functions $S$ in (\ref{eq:prof}) depend on the polynomial $p(z)$ in (\ref{eq:defenge}),
and even when $\efold_a=\efold_b$ and the same polynomial $p(z)$ is used,
the field growth and decay will not be perfectly symmetric with respect
to the center of the magnet.
This is due to the even powers in $p(z)$.
In order that the rise and fall
of $\mathcal{S}(z)$ be symmetric we must change
the sign of all the even terms between the entrance polynomial $p_a(z)$ and the exit
polynomial $p_b(z)$.
The hyperbolic tangent polynomial $p(z)=z$ contains no even terms and requires no modification.

In order to provide sufficient room for the fringe field decay,
we must increase the physical length of the ``magnet",
at least a few units of $\efold$ on each end.
Use of the hyperbolic tangent function ($p(z)=z$) preserves the
integrated magnet strength, but the general Enge function does not
and may require some adjustment to the overall strength $B_0$.

We require the magnet to have mid-plane symmetry, or that
\begin{eqnarray}
b_x(x,y,z) & = & -b_x(x,-y,z) , \label{eq:sym1}\\
b_y(x,y,z) & = &  b_y(x,-y,z) , \label{eq:sym2}\\
b_z(x,y,z) & = & -b_z(x,-y,z) \label{eq:sym3}.
\end{eqnarray}
If we consider $T(x)$ and $S(z)$ as known functions, then the symmetry constraints (\ref{eq:sym1}-\ref{eq:sym3})
plus Maxwell's equations determine the field everywhere, and can be obtained using a power series expansion.
The magnetic potential $\psi$ can be written as the power series
\begin{equation}
\psi = y f_0(x,z) + \frac{y^3}{3!}f_1(x,z) + \frac{y^5}{5!}f_2(x,z) + \ldots
\label{eq:PS0}
\end{equation}
where $\v{B}=\grad\psi$ and $\grad^2 \psi=0$.
Here the symmetry dictates that $\psi$ must be an odd function of $y$.
The given transverse profile implies $f_0(x,z)=T(x)S(z)$ and in order that $\grad^2 \psi=0$,
\begin{equation}
f_n(x,z) = (-1)^{n} \sum_{i=0}^n \binom{n}{i} T^{(2i)}(x) S^{(2n-2i)}(z) .
\end{equation}
In the special case with a linear gradient in the field, $T(x)=B_0+Gx$, this simplifies to
\begin{equation}
f_n(x,z) = (-1)^{n} (B_0+Gx) S^{(2n)}(z)
\end{equation}

The power series (\ref{eq:PS0}) is valid for any $x$ and $z$,
but converges only when $y$ is small enough.
For the $\tanh$ profile function, $p(z)=z$, we can obtain a bound on the radius of convergence.
As shown in Appendix~A, the field is guaranteed to converge for $|y|<R$, where
\begin{equation}
R = \frac{\pi}{2} \efold .
\label{eq:rval}
\end{equation}
In the center of the fringe field region,
$R$ is equal to the radius of convergence for the power series (\ref{eq:PS0}).
At any other value of $x$ and $z$, it is a lower bound on the radius on convergence.

If we allow the fringe fields to be arbitrarily short, $\efold\rightarrow 0$,
the field diverges for arbitrarily small $y$.
The reason for this divergence is that the rapid growth of the derivatives of the
longitudinal profile function $S^{(2n})(z)$.
It is not physically realistic to allow the fringe field to decay over too short a length.
The length of fringe field decay should instead be set according to the magnet aperture.

Walstrom \cite{Walstrom2004} has also computed the radius of convergence for magnetic field power series.
He uses the on-axis generalized gradients, which are considered as complex-valued functions.
The radius of convergence at a particular location can then be determined by the nearest singularity
in the complex plane.
However, he does not calculate this radius of convergence for the case of a bend magnet with mid-plane
symmetry.


Venturini \cite{Venturini1999} has calculated higher order aberrations associated
with hard-edge fringe field kicks in quadrupole magnets.
He found that magnets with shorter fringe fields have larger high-order aberrations.
In our view this can be explained by our calculation that the magnetic field
power series diverges for all $y>0$ as the fringe field length goes to zero.

Instead of assuming a magnet with mid-plane symmetry,
we could have used a conventional multipole field with the same longitudinal profile function $S(z)$.
The field is then a function of the radial coordinate $r$, the angle $\theta$ and the longitudinal variable $z$.
We have repeated the same power series analysis, and find again that the magnetic potential power series is guaranteed
to converge for $|r|<R$, where $R$ is again given by (\ref{eq:rval}).
This multipole field may not converge for large values of the transverse variable $x$,
whereas with the magnet with mid-plane symmetry the convergence of the power series
(\ref{eq:PS0}) depends only upon $y$.
Since transverse excursions of 0.2 m are present in our model FFAG accelerator,
for soft-edge fringes we use field expansions with mid-plane symmetry rather
than a quadrupole field.

\section{Modeling in PTC}

For modeling FFAGs, we use the Polymorphic Tracking Code (PTC),
a beam tracking code developed by
\'{E}tienne Forest \cite{Forest06,Machida2001}.
For our purposes, the most important--indeed, essential---reasons for
using PTC are its ability to place magnets in arbitrary locations
(needed for the large transverse offsets of the quadrupoles), its 
correct handling of the large energy range in an FFAG, and its ability
to integrate through arbitrary magnetic fields.
This last is done using an element of type \textit{arbitrary}.

Before any particle tracking, each \textit{arbitrary} magnet is analyzed.
Each component of the magnetic field in each transverse slice is fit
by a bivariate polynomial $\sum c_{ij} x^i y^j$, 
with $0 \le i+j\le 8$.
The fitting is done using a least-squares minimization.
The magnetic field given by the potential (\ref{eq:PS0}) is a polynomial in $x$ and $y$, but not $z$.
The fitting is necessary because the transverse slices are rotated from this orientation.
To track a particle through this element, PTC then integrates using these pre-calculated
polynomial fields by means of a Lorentz kick.
Here this integration scheme is not symplectic,
but it is sufficient for short term tracking ($<1000$ turns).

In element of type \textit{arbitrary} the fringe field region is taken into account directly through the
rise in the magnitude of the magnetic field, and not using a delta function kick.
This is more accurate, but presents several difficulties.
First, the integration step must be small in the fringe field,
where the field is increasing rapidly.
Second, the magnet must be longer to accommodate the fringe fields.
The combined function bends are separated by a short drift of only 7.5~cm,
so that their fringe fields overlap.
We have modeled this section by making the magnets 15~cm longer on each end,
as can be seen in Figure~\ref{figtrack0}.
In order to patch the various overlapping components together, PTC automatically drifts the particles
backwards to the beginning of the next component.
These backward tracks can be seen in Figure~\ref{figtrack0}.
This method of integrating between overlapping components is justified
by the fact that the field is small in the region of overlap.

In all cases we use a field profile $S(z)$ with a hyperbolic tangent profile, $p(z)=z$.
The decay length $\efold$ was set to 5~cm, which is similar to the aperture.
We use an integration step of 1~cm within elements of type \textit{arbitrary}.

An alternate approach would be to consider the two bend magnets as one long magnet.
A 3D magnet code (such as TOSCA) can be used to generate the
magnetic field of both magnets together as a single beamline element.
We did not do this because it would entail additional complexity,
for one thing the two magnets are rotated with respect to one another.
Our model represents a compromise between a standard tracking code using hard-edge fringe kicks,
and a model which tracks through many magnets using the full 3D field.
We retain the convenience of modeling the beamline as a simple sequence of elements
(DL, BD, DS, BF) while reproducing more realistic fringe fields.


\section{Results}

\begin{figure}[htb]
\center
\includegraphics{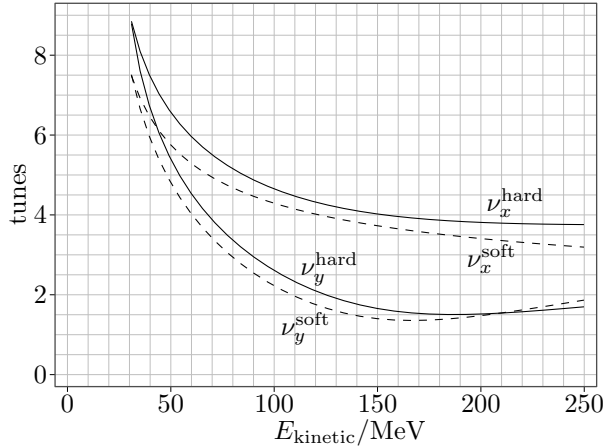}
\caption{Machine tune versus kinetic energy.
The solid lines are for hard-edged magnets, the dashed lines for soft-edge (field mapped) magnets.}
\label{figtune0}
\end{figure}

We ran PTC both with the special field-mapped magnets, and also with conventional quadrupoles
with a hard-edge fringe field entrance and exit kick.
These magnet types differ primarily in their treatment of the fringe field.
Figure~\ref{figtune0} shows the horizontal and vertical tunes for both magnet types.
The tunes are similar but differ in key areas.
In particular, the vertical and horizontal tunes are over 8 after injection for the conventional quadrupoles,
but both are slightly under 8 for the field-mapped magnets.
A machine tune of 8 corresponds to a single-period tune of 1/3,
because we have a 24 cell lattice.

\begin{figure}[htb]
\center
\includegraphics{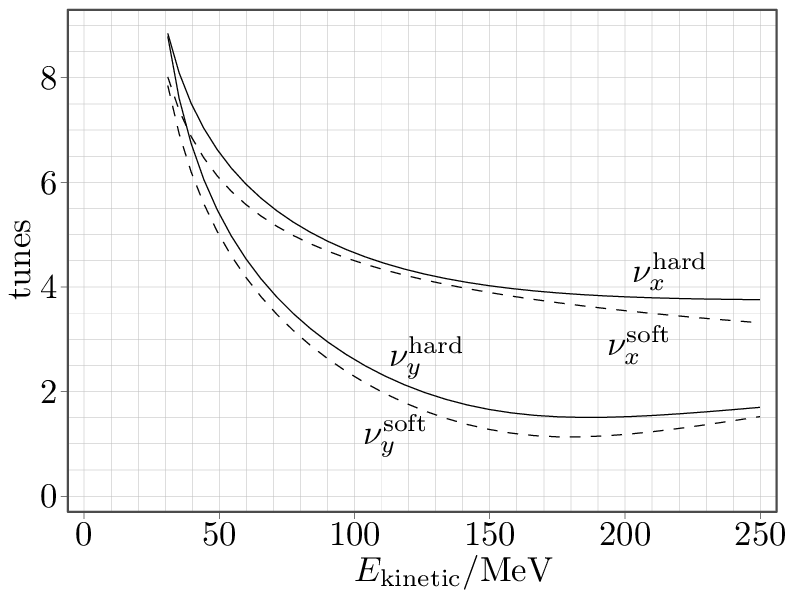}
\caption{Machine tunes versus kinetic energy.  Solid lines unchanged,
dashed lines are for soft-edge
quads with increased strength $G$ at 8.4 and -13.2 T/m.}
\label{figtune1}
\end{figure}

\begin{figure}[htb]
\center
\includegraphics{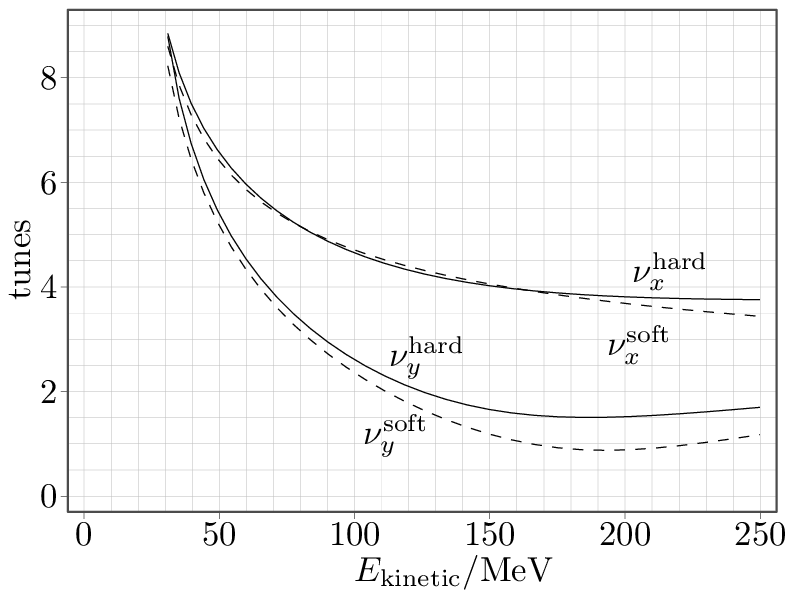}
\caption{Machine tunes versus kinetic energy.  Solid lines unchanged,
dashed lines are for soft-edge
quads with increased strength $G$ at 8.8 and -13.6 T/m.}
\label{figtune2}
\end{figure}

If we increase the strength of the quadrupoles, or equivalently increase the gradient of the field $G$,
the tune diagrams become more similar to those for hard-edge magnets.
Figure~\ref{figtune1} shows the result when $G$ is increased to 8.4 and -13.2 T/m,
and Figure~\ref{figtune2} when $G$ is increased to 8.8 and -13.6 T/m.
For low energies the tunes are very close to the hard-edge case, and in particular
the tune 8 resonance must again be crossed.
There are still differences in the tunes at higher energies.

\begin{figure*}[htb]
\includegraphics{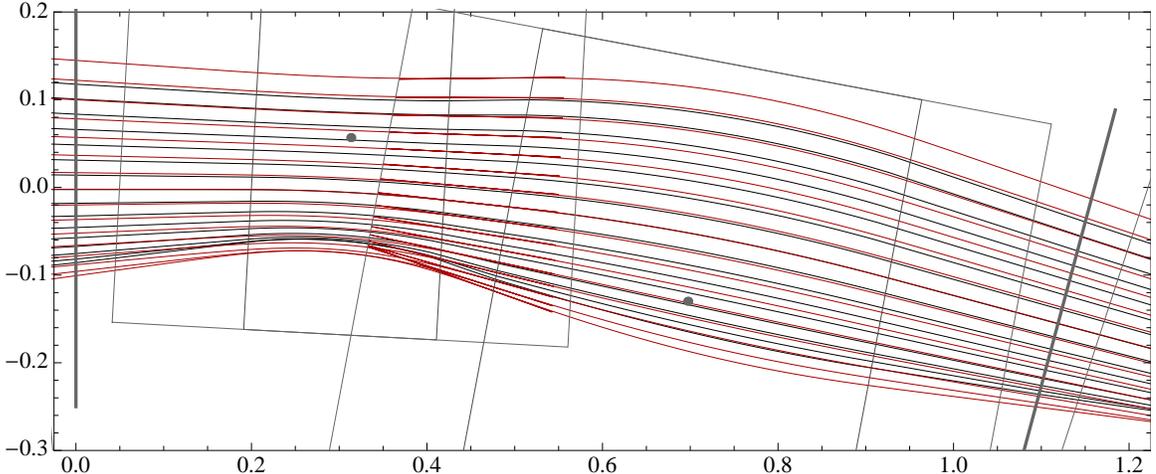}
\caption{Top view of a single period showing closed orbits for hard-edge magnets
(black, same as Figure~\ref{figtr1}) and soft-edge magnets (red) with $G$ at 8.0 and -12.8 T/m. The scale is in meters.}
\label{figtrack0}
\end{figure*}

Figure~\ref{figtrack0} shows closed orbits for both conventional hard-edge fringes
and using the new soft-edge magnets.
We can see that the horizontal focusing is decreased by the soft-edge fringes,
so the transverse excursion of the orbits is greater.

\begin{figure*}[htb]
\includegraphics{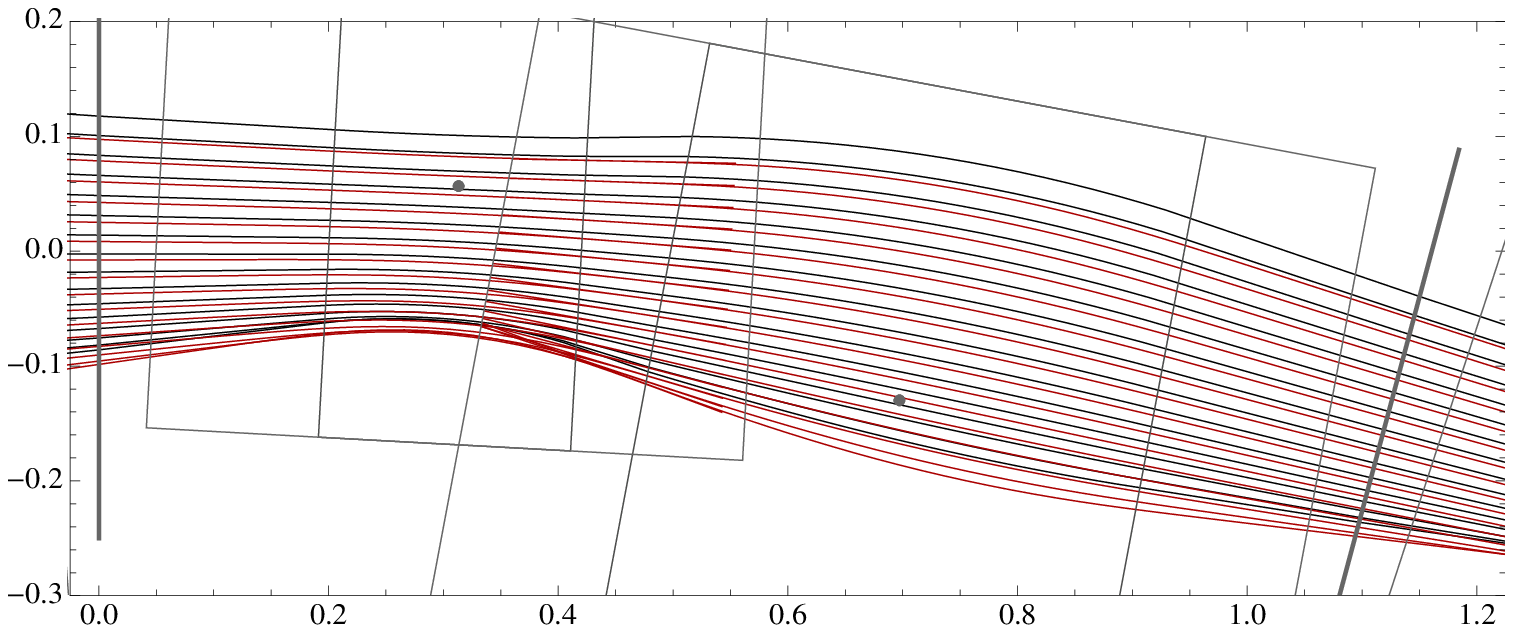}
\caption{Top view of a single period showing closed orbits for hard-edge magnets
(black, same as Figure~\ref{figtr1}) and soft-edge magnets (red) with $G$ at 8.8 and -13.6 T/m. The scale is in meters.}
\label{figtrack2}
\end{figure*}

Figure~\ref{figtrack2} shows closed orbits for the magnets with the largest gradient (8.8 and -13.6 T/m),
compared with the closed orbits of the original hard-edge fringes (same as in Figure~\ref{figtrack0}).
We can see that the horizontal focusing now appears similar,
although the soft-edge orbits are displaced inwards, particularly at higher energies.

We now take 1000 protons and accelerate them from 31 to 250~MeV using PTC.
The starting emittance was $16\pi$ mm-mrad in each plane.
The particles were uniformly distributed inside a phase space ellipse.
In order to reduce the instabilities due to the many resonances that must be crossed,
the acceleration needs to take place as rapidly as possible.
For this reason we place an RF cavity in each of the long drifts.
The 24 RF cavities add 350 KeV per turn, so acceleration takes 626 turns.
The current in this machine is small enough that we do not take
into account space charge.

\begin{figure}[htb]
\center
\includegraphics[width=3.4in]{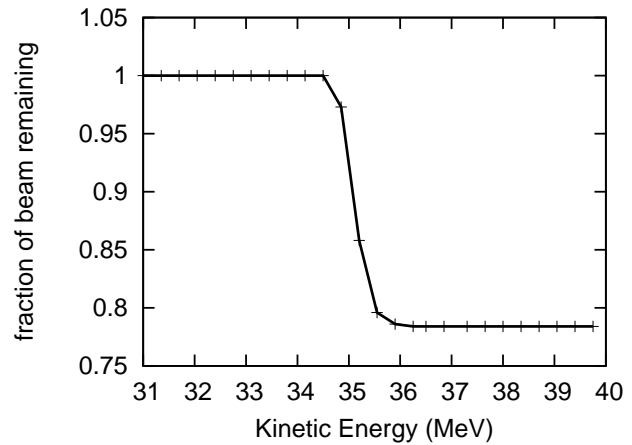}
\caption{Fraction of the beam lost by energy using hard-edge magnets.}
\label{figloss}
\end{figure}

Using conventional hard-edge fringe field magnets,
21\% of the beam is lost in the first 20 turns.
This occurs at the crossing of the 1/3 tune (Figure~\ref{figloss}).
No other resonance gives further loss of the beam.
Using the special field-mapped magnets, none of the beam is lost.
This is not surprising because the tunes no longer cross the $1/3$ resonance.

If we increase the quadrupole gradient of the magnets,
the tunes again cross the $1/3$ resonance,
but a 21\% beam loss is not seen using the soft-edge fringes.
With the magnets of Figure~\ref{figtune1} only 3.4\% of the particles are lost and with
Figure~\ref{figtune2} only 5.6\% are lost.

The hard-edge fringe field kick derived in \cite{Forest88} and used for quadrupoles in PTC
assumes that the fringe field has quadrupole symmetry.
A subtle point is that
since our field is assumed to have mid-plane symmetry,
the quadrupole fringe field kick may not be accurate for
fringe fields with mid-plane symmetry.
It may be that some of the particle loss differences between hard-edge and soft-edge magnets
is due to this problem.
Ideally, one would like to repeat the calculation in \cite{Forest88} for a magnet with mid-plane
symmetry, and compare results using this fringe field kick to the soft-edge fringe model,
but this is not a simple calculation and we have not done this.
The fact that the closed orbits for hard and soft-edge magnets in Figure~\ref{figtrack0}
are so close to one another suggests to us that our hard-edge fringe field kicks
are reasonably accurate.

Our results indicate that modeling fringe fields correctly is critical to
predict the performance of non-scaling FFAGs.
Although particle tracking with the new field-mapped, soft-edge magnets is about 10 times
slower than using conventional hard-edge fringe model,
the additional computational time far outweighs any later re-engineering cost after a machine is built.
In addition, the fact that we are only tracking particles through 626 turns, rather than
millions of turns as in a storage ring, means that the tracking time for 1000 protons goes
from seconds to minutes, in other words is still easily handled.


\section{Summary}
We have used the tracking code PTC to track protons through the acceleration phase
of a non-scaling FFAG from 31 to 250~MeV.
We have compared the use of a conventional, hard-edge fringe model with a more accurate,
soft-edge fringe model.
With an emittance of $16\pi$ mm-mrad,
the hard-edge fringe model predicts a 21\% particle loss as the beam moves
through the 1/3 tune resonance.
In contrast the more accurate fringe field model predicts only a 6\% particle loss.


A future paper will study the effect of magnet misalignments for this machine \cite{Abell2011}.

\section*{Acknowledgment}
This work was supported in part by the U.S. Department of Energy, Office of Science,
Office of Nuclear Physics under SBIR Grant No.\,DE-FG02-06ER84508.

\appendix
\section{Enge function derivatives and radius of convergence}
\label{S:Enge}
Differentiation of the Enge function (\ref{eq:defenge}) is straightforward
but results in formulas of rapidly increasing complexity.
The results are easier to comprehend if we interpret the Enge function as a composition of two functions,
$S(z;\efold) = f\circ g(z)$ where
\begin{eqnarray}
f(z) & = & \frac{1}{2}\left(1-\tanh z\right) \mbox{ and} \\
g(z) & = & p(z/\efold) .
\end{eqnarray}


\begin{figure*}[!t]
\normalsize
\begin{equation}
\frac{\ud^n}{\ud z^n} f\circ g(z) = \sum_{k=1}^n f^{(k)}\circ g(z)\sum_{part}\diB(n; a_1, a_2, \ldots, a_n)
(g'(z))^{a_1}(g''(z))^{a_2} \ldots (g^{(n)}(z))^{a_n} \label{eq:diBruno}
\end{equation}
\begin{equation}
\frac{\ud^n}{\ud z^n} f\circ g(z) = \sum_{k=1}^n f^{(k)}\circ g(z)
\sum_{part}
\frac{n!}{a_1! a_2! \cdots a_n!}
\prod_{i=1}^n\left(\frac{g^{(i)}(z)}{i!}\right)^{a_i} \label{eq:diBruno1}
\end{equation}
\hrulefill
\vspace*{4pt}
\end{figure*}
Derivatives of $S$ can be found using \textit{Fa\`{a} di Bruno's formula} \cite{AS}
for differentiation of a composition of two functions
where the second sum in (\ref{eq:diBruno}) or (\ref{eq:diBruno1}) is over
the partitions a set of $n=a_1+2a_2+ \ldots + na_n$
different objects into $a_k$ subsets containing $k$ objects for $k=1,2,\dots, n$,
and $k=a_1+a_2+ \ldots +a_n$.
The coefficient $\diB(n; a_1, a_2, \ldots, a_n)$ is the number of such partitions.
These are related to multinomial coefficients and are defined by
\begin{equation}
\diB(n; a_1, a_2, \ldots, a_n) = \prod_{i=1}^n \frac{i}{(i!)^{a_i} a_i!} .
\end{equation}




For (\ref{eq:diBruno}) and (\ref{eq:diBruno1}) we need derivatives of
$\tanh(z)$ to arbitrary order.
We find that
\begin{equation}
\frac{\ud^n \tanh z}{\ud z^n} = q_n(\tanh z)
\label{eq:dtanh}
\end{equation}
where $q_n$ is a polynomial of degree $n+1$.
These polynomials are defined by $q_0=z$ and the recurrence relation
\begin{equation}
q_{n+1}(z) = (1-z^2)\frac{\ud q_n}{\ud z}
\end{equation}
In \cite{Boyadzhiev},
it is shown that these polynomials can be written in terms of Stirling Numbers of the second kind,
$S_2(n,k)$.
\begin{equation}
q_n(z) = (-2)^n (z+1) \sum_{k=0}^n S_2(n,k) k! \left(\frac{z-1}{2}\right)^k
\label{eq:stirling}
\end{equation}

\begin{table*}[htb]
  \caption{\label{table1}The first 12 polynomials $q_n(z)$.}
  \centering
  \begin{tabular}{cl}
    $n$ & $q_n(z)$ \\ \hline
     0 & $z$ \\
     1 & $1 -z^2$ \\
     2 & $-2z(1-z^2)$ \\
     3 & $-2(1-z^2)(1-3z^2)$ \\
     4 & $8z(1-z^2)(2-3z^2)$ \\
     5 & $8(1-z^2)(2 -15z^2 +15z^4)$ \\
     6 & $-16z(1-z^2)(17 -60z^2 +45z^4)$ \\
     7 & $-16(1-z^2)(17 -231z^2 +525z^4 -315z^6)$ \\
     8 & $128z(1-z^2)(62 -378z^2 +630z^4 -315z^6)$ \\
     9 & $128(1-z^2)(62 -1320z^2 +5040z^4 -6615z^6 +2835z^8)$ \\
    10 & $-256z(1-z^2)(1382 -12720z^2 +34965z^4 -37800z^6 +14175z^8)$ \\
    11 & $-256(1-z^2)(1382 -42306z^2 +238425z^4 -509355z^6 +467775z^8 -155925z^{10})$ \\
    12 & $1024z(1-z^2)(21844  -280731z^2 +1121670z^4 -1954260z^6 +1559250z^8 -467775z^{10})$ \\
    \hline
  \end{tabular}
\end{table*}

\begin{figure}[htb]
\includegraphics[scale=1.3]{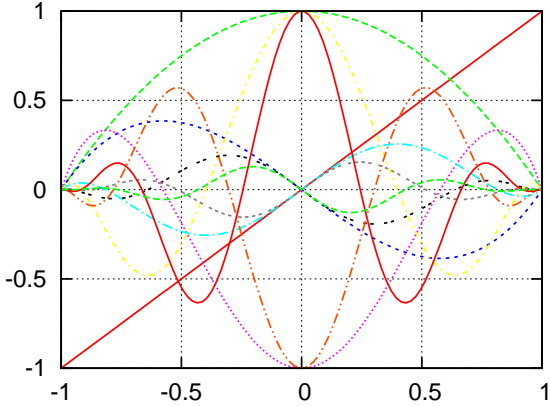}
\caption{The polynomials $q_0(z)$ to $q_{10}(z)$
normalized by $t_{\lceil n/2\rceil}$.}
\label{figenge}
\end{figure}

Table~\ref{table1} gives the polynomials $q_n$ up to $n=12$,
and Figure~\ref{figenge} shows them graphically up to $n=10$.
Table~\ref{table2} gives the coefficients needed in Fa\`{a} di Bruno's formula (\ref{eq:diBruno})
to calculate the 5th derivative of $f\circ g(z)$.
As an example,
using the results of Tables~\ref{table1} and \ref{table2},
we calculate the 5th derivative of the Enge function $S(z;\efold)$ as
\begin{eqnarray}
\frac{\ud^5 S}{\ud z^5}  & = & -\frac{1}{2\efold^5}\left[p^{(5)}q_1 + (5p^{(4)}p'+10p^{(3)}p'')q_2 \right. \nonumber \\
 & & + (10p^{(3)}p'p'+15p''p''p')q_3 \nonumber \\
  & & \left. + 10p''(p')^3 q_4 + (p')^5 q_5 \right]
\label{eq:d5Enge}
\end{eqnarray}
where derivatives of the polynomials $p$ are evaluated at $z/\efold$, and the polynomials $q_k$
are evaluated at $\tanh(g(z))=\tanh(p(z/\efold))$.
In the special case of the $\tanh$ function, where $p(z)=z$, all the terms in (\ref{eq:d5Enge})
are zero except for the last one and we have
\begin{equation}
\frac{\ud^5 \tanh(z/\efold)}{\ud z^5} = \frac{q_5(\tanh(z/\efold))}{\efold^5} ,
\end{equation}
which is the same as (\ref{eq:dtanh}) for $n=5$.

\begin{table}[htb]
  \caption{\label{table2}
    The seven terms in $S^{(5)}$.}
  \centering
  \begin{tabular}{cllc}
    $k$ & \text{partition of 5} & \text{multin. coeff} & \text{term} \\
    \hline
    1 & 5         & $\diB(\mbox{\footnotesize 5;0,0,0,0,1})$ &  $p^{(5)} q_1$ \\
    2 & 4+1       & $\diB(\mbox{\footnotesize 5;1,0,0,1,0})$ & $5p^{(4)}p' q_2$ \\
    2 & 3+2       & $\diB(\mbox{\footnotesize 5;0,1,1,0,0})$ & $10p^{(3)}p'' q_2$ \\
    3 & 3+1+1     & $\diB(\mbox{\footnotesize 5;2,0,1,0,0})$ & $10p^{(3)}(p')^2 q_3$ \\
    3 & 2+2+1     & $\diB(\mbox{\footnotesize 5;1,2,0,0,0})$ & $15(p'')^2 p' q_3$ \\
    4 & 2+1+1+1   & $\diB(\mbox{\footnotesize 5;3,1,0,0,0})$ & $10p''(p')^3 q_4$ \\
    5 & 1+1+1+1+1 & $\diB(\mbox{\footnotesize 5;5,0,0,0,0})$ & $(p')^5 q_5$ \\
    \hline
  \end{tabular}
\end{table}

The absolute value of the lowest order coefficient in $q_n(z)$
(the constant or linear term, whichever is nonzero) generates the sequence
$1$, $1$, $2$, $2$, $16$, $16$, $272$, $\ldots$.
The sequence $1$, $2$, $16$, $272$, $7936$, $353792$, $\ldots$ is sequence A182
in OEIS \cite{OEIS}, known as the ``tangent numbers" $t_n$.
These are Taylor series coefficients of the
$\tanh(z)$ function expanded about $z=0$,
and may also be written in terms of Bernoulli numbers \cite{OEIS}.
In Figure~\ref{figenge} the polynomials $q_n(z)$ are divided by the
normalizing factor $t_{\lceil n/2\rceil}$, and it is clear that
\begin{equation}
|q_n(z)|\le t_{\lceil n/2\rceil} \mbox{ for } |z|\le 1 ,
\label{eq:ceil}
\end{equation}
at least for $n\le 12$.
For odd $n$, (\ref{eq:ceil}) is equivalent to the statement
that the maximum magnitude of the $n$'th derivative of $\tanh(z)$
occurs at $z=0$.
This can be proven by writing the Taylor expansion of the $n$'th derivative of $\tanh(z)$ about $z=0$
and applying the Alternating Series Estimation Theorem.

Since the lowest order coefficient of $q_n(z)$ is $ t_{\lceil n/2\rceil}$,
in the center of the fringe field we can say:
\begin{equation}
q_n(0) = (-1)^{(n-1)/2}
\left\{ \begin{array}{lc}
t_{(n+1)/2} & n \mbox{ odd}, \\
0 & n \mbox{ even},
\end{array} \right.
\label{eq:qn0}
\end{equation}
and

\begin{equation}
\frac{\ud q_n}{\ud z}(0) = (-1)^{n/2}
\left\{ \begin{array}{lc}
0 & n \mbox{ odd}, \\
t_{n/2} & n \mbox{ even}.
\end{array} \right.
\label{eq:qnp0}
\end{equation}

The asymptotic form for the tangent numbers is given by \cite{OEIS}
\begin{equation}
t_n \cong 2\left(\frac{2}{\pi}\right)^{2n} (2n-1)! .
\label{eq:A182}
\end{equation}
This formula is quite accurate even for small $n$---for example it predicts $t_4\cong 271.958$,
when $t_4=272$.
We can use this asymptotic form to obtain a bound on the radius of convergence for the
magnetic field power series (\ref{eq:PS0}) for the $\tanh$ longitudinal profile
and linear transverse field gradient (\ref{eq:lingrad}).
If the power series terms in (\ref{eq:PS0}) are $c_n$, i.e. $\psi=\sum c_n y^n$,
then $c_n=0$ for even $n$ and
Equation (\ref{eq:A182}) gives the bound
\begin{equation}
|c_n| \le \frac{|B_0+Gx|}{n(n-1)} \left(\frac{2}{\pi\efold}\right)^{n-1} \mbox{ for } n>1 .
\label{eq:abound}
\end{equation}
The radius of convergence of the power series with coefficients $c_n$ is given by
$1/R = \limsup_{n\rightarrow\infty} \sqrt[n]{\left| c_n \right|}$,
which leads directly to the bound
\begin{equation}
R = \left( \limsup_{n\rightarrow\infty} \sqrt[n]{|c_n|}\right)^{-1} \ge \frac{\pi\efold}{2}
\end{equation}
Thus the power series (\ref{eq:PS0}) converges for $|y|<\pi\efold/2$
(for any value of $x$ and $z$).


\section*{Acknowledgment}
This work was supported in part by the U.S. Department of Energy, Office of Science,
Office of Nuclear Physics under SBIR Grant No.\,DE-FG02-06ER84508.

\end{document}